\documentclass[twocolumn,aps,prd,epsf]{revtex4}
\usepackage{epsfig}

\begin{document}

\title{The family of strange multiquarks related to the $D_s(2317)$ and $D_s(2457)$}
\author{P. Bicudo}
\email{bicudo@ist.utl.pt}
\affiliation{
Dep. F\'{\i}sica and CFIF, Instituto Superior T\'ecnico, 
Av. Rovisco Pais
1049-001 Lisboa, Portugal}
\begin{abstract}
I study the $D_s(2317)$ and $D_s(2457)$ discovered at BABAR, CLEO and BELLE, and find that 
they belong to a class of strange $S=-1$ tetraquarks and pentaquarks, which is equivalent to the 
class of kaonic molecules bound by short range attraction. 
In this class of hadrons a kaon is strongly trapped by a s-wave meson or baryon.
To describe this class of multiquarks the Resonating Group Method is applied
to a standard quark model with chiral symmetry breaking, 
and the short range kaon-meson(baryon) interactions are extracted. 
A criterion is derived to classify the attractive channels. 
I conclude that the mesons $B_s^{(0+)}, \, B_s^{(1+)}$ , 
and the  baryons $\Omega_{cc}, \, \Omega_{cb}, \, \Omega_{bb}$ clearly belong to the new hadronic 
class of the $D_s(2317)$ and $D_s(2457)$. 
The hadrons $f_0(980), \, \Lambda, \, \Sigma_c, \, \Sigma_b$ possibly belong to a related family. 
\end{abstract}
\maketitle

\section{Introduction}

\par
Recently new narrow scalar resonances $D_s(2317)$ and $D_s(2457)$ were 
discovered at BABAR  
\cite{Babar}, 
Cleo 
\cite{Cleo}
and Belle 
\cite{Belle}.
They are not expected to be standard hadrons because their masses are
close to hundred MeV smaller than the ones predicted in quark models for 
positive parity $D_s$ quark-antiquark mesons 
\cite{Isgur}. For instance the experimental candidate to the
first quark-antiquark $D_s$ meson is the $D_{s1}(2536)$
\cite{PDB}.
This is confirmed by lattice simulations 
\cite{Bali}
for quark-antiquark mesons. The existence of non-mesonic and non-hadronic 
multiquarks has been suggested form the onset of the quark model
\cite{Jaffe,Strottman}.
These positive parity $D_s$ resonances are interpreted by Beveren and Rupp 
\cite{Rupp2}
as poles in the S Matrix of the coupled channels of mesons and meson pairs.
They were also predicted by Nowak, Rho and Zahed 
\cite{Nowak,Bardeen}
as chiral partners of the well known negative parity pseudoscalar $D_s(1968)$ 
and $D_s(2112)$.
They are as well interpreted as K-D molecules by Barnes, Close and Lipkin
\cite{Barnes1},
similar to the Deusons anticipated by Tornqvist
\cite{Tornqvist}.
In the same way I find here that the new $D_s$ resonances can be understood as 
tetraquarks, or equivalently as s-wave $D$-$K$ molecules bound by the short range 
attraction.

\par
The experimental discovery of these hadron also suggests the existence of
a new class of multiquarks.
In this paper I study the family of all possible narrow tetraquark 
and pentaquark resonances 
where the quark $s$, or the $S=-1$ Kaon play a crucial role. 
The strangeness $S=1$ pentaquark $\theta^+$ is more difficult
to bind and was recently addressed with the same techniques of
this paper in reference
\cite{Bicudo5}.
A chiral invariant framework 
\cite{Bicudo3,Bicudo1,Bicudo2} is used to compute  microscopically, at the 
quark level, the masses of this new class of hadrons.
The resulting mechanism which provides the binding of the $S=-1$ tetraquarks 
and pentaquarks is equivalent to the short range attraction of hadrons. 

\par
Here a standard Quark Model (QM) Hamiltonian is assumed, 
\begin{equation}
H= \sum_i T_i + \sum_{i<j} V_{ij} +\sum_{i \bar j} A_{i \bar j} \,
\label{Hamiltonian}
\end{equation}
where each quark or antiquark has a kinetic energy $T_i$ with a
constituent quark mass, and the colour dependent two-body
interaction $V_{ij}$ includes the standard QM confining and 
hyperfine terms,
\begin{equation}
V_{ij}= \frac{-3}{16} \vec \lambda_i  \cdot  \vec \lambda_j
\left[V_{conf}(r) + V_{hyp} (r) { \vec S_i } \cdot { \vec S_j }
\right] \ . \label{potential}
\end{equation}
Moreover the Hamiltonian
includes a quark-antiquark annihilation term $A_{i \bar j}$
which is the result of spontaneous chiral symmetry breaking. 

\par
This paper is organised in sections. In Section II the QM is reviewed,
together with the Resonating Group Method (RGM)
\cite{Wheeler}
which is adequate to study multiquark states, where several quarks overlap. 
The RGM, together with chiral symmetry, produces short range
hadron-hadron potentials, which can be either repulsive (hardr core repulsion)
or attractive. 
In Section III a criterion is derived to discriminate which systems
bind and which are unbound. This criterion is applied to find, 
among the s-wave hadrons, the candidates to trap a kaon. 
In section IV the binding energy is computed for the selected 
positive parity mesons $f_0(980)^{(0+)},  D_s^{(0+)}, \, D_s^{(1+)}, 
\, B_s^{(0+)}, \, B_s^{(1+)}$ and negative parity baryons 
$\Lambda, \, \Xi_c, \, \Xi_b,  \, \Omega_{cc}, 
\, \Omega_{cb}, \, \Omega_{bb}$.
Finally the results are presented and discussed in Section V.

\par

\section{Studying multiquarks with the RGM}

For the purpose of this paper the details of potential
(\ref{Hamiltonian}) are unimportant, only its matrix elements matter. 
The hadron spectrum constrains the hyperfine potential,
\begin{equation}
\langle V_{hyp} \rangle \simeq \frac{4}{3} \left( M_\Delta-M_N
\right) \label{hyperfine}
\end{equation}
The quark-antiquark annihilation potential $A_{i \bar j}$ is 
also constrained when the quark model produces spontaneous chiral 
symmetry breaking
\cite{Bicudo3,Bicudo4}.
The annihilation potential $A$ is present in the $\pi$
Salpeter equation,
\begin{equation}
\left[
\begin{array}{cc}
2 T + V & A \\
A & 2T +V
\end{array}
\right]
\left(
\begin{array}{c}
\phi^+ \\
\phi^-
\end{array}
\right) =
M_\pi
\left(
\begin{array}{c}
\phi^+ \\
-\phi^-
\end{array}
\right)
\label{pion BS}
\end{equation}
where the $\pi$ is the only hadron with a large negative energy
wave-function, $\phi^- \simeq \phi^+$.  In eq. (\ref{pion BS}) the
annihilation potential $A$ cancels most of the kinetic energy and
confining potential $2T+V$. This is the reason why the pion has a very
small mass. From the hadron spectrum and using eq. (\ref{pion BS})
the matrix elements of the annihilation potential are determined,
\begin{eqnarray}
\langle 2T+V \rangle_{S=0} &\simeq& {2 \over 3} (2M_N-M_\Delta)
\nonumber \\
\Rightarrow \langle A \rangle_{S=0} &\simeq&- {2 \over 3} 
(2M_N-M_\Delta)
\ ,
\label{sum rules}
\end{eqnarray}
where this result is correct for the annihilation of $u$ or $d$ quarks.
When a strange quark is present, the corresponding matrix element 
is smaller by a factor $\sigma$ which is a power of the constituent 
quark mass ratio $M_{u,d} /M_s$

\par
%
%
%
\begin{figure}[t]
%
\begin{picture}(200,80)(0,0)
\put(20,5){
\begin{picture}(120,30)(0,0)
\put(20,0){
\begin{picture}(100,100)(0,0)
\put(0,20){\vector(1,0){100}}
\put(50,5){$\vec \rho_A= {\vec r_1 -\vec r_2 \over \sqrt{2}}$}
\put(60,60){\vector(-1,-4){10}}
\put(65,41){$\vec \lambda_{A,B}={\vec r_1 +\vec r_2 -\vec r_3 -\vec r_4 \over 2}$}
\put(10,60){\vector(1,0){100}}
\put(60,70){$\vec \rho_{B}={\vec r_3 -\vec r_4 \over \sqrt{2}}$}
\end{picture}}
\put(20,0){
\begin{picture}(100,100)(0,0)
\put(-15,18){$\bar q_2$}
\put(0,20){\circle*{5}}
\put(110,18){$q_1$}
\put(102,20){\circle*{5}}
\put(-5,58){$\bar q_4$}
\put(10,60){\circle*{5}}
\put(120,58){$q_3$}
\put(112,60){\circle*{5}}
\end{picture}}
\end{picture}}
\end{picture}
%
\caption{
The Jacobi coordinates of the incoming four quark wave-function
$\phi_A(\rho_A) \, \phi_B(\rho_B) \, \psi(\lambda_{A,B})$.
}
\label{RGM coordinates}
\end{figure}
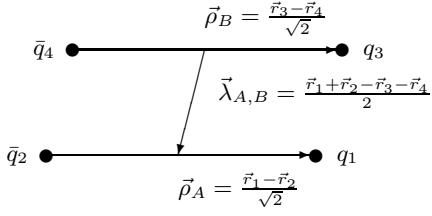
The QM of eq. (\ref{Hamiltonian}) 
reproduces the meson and baryon spectrum with quark and antiquark
bound-states (from the heavy quarkonium to the light pion mass).
Moreover the RGM was first used in hadronic physics by Ribeiro 
\cite{Ribeiro} 
to show that in exotic
hadron-hadron scattering, the quark-quark potential together with
the Pauli repulsion of quarks produces a repulsive short range
interaction. For instance this explains the $N - N$ hard core
repulsion, preventing the collapse of nuclei. 
This $N - N$ hard core repulsion is supposed to also occur
in several hadron-hadron interactions, and this explains why 
many multiquarks systems are not stable. However Deus and Ribeiro
\cite{Deus} 
used the same RGM to show that, in non-exotic channels, the 
quark-antiquark annihilation could produce a short core attraction. 
Recently it was shown that in the particular case of the low energy 
$\pi-\pi$ system in the chiral limit, the short range attraction and 
repulsion exactly cancel
\cite{Bicudo1},
resulting in a Adler Zero and the Weinberg theorem. 
Addressing a tetraquark system with the $\pi-\pi$ quantum numbers,
it was shown that the QM also fully complies with the chiral 
symmetry, including the PCAC theorems
\cite{Bicudo1}.
Therefore the QM is adequate to address the new $D_s$ hadrons, which 
were predicted by Nowak, Rho and Zahed in an effective chiral model. 
In this paper the QM and the RGM are applied to $S=-1$ multiquarks.

\par
The RGM 
\cite{Wheeler}
computes the effective multiquark energy 
using the matrix elements of the microscopic quark-quark interactions. 
Any multiquark state can be decomposed in combinations of simpler 
colour singlets, the baryons and mesons. This can be illustrated with
the colour structure of a multiquark with two quark-antiquark pairs,
$q_1, \, \bar q _2, \, q_3, \, \bar q_4$. 
Assuming that this tetraquark is a colour singlet, each quark-antiquark 
pair can either be a colour singlet or a colour octet. Nevertheless
the octet-octet state can be described with an exchange operator
and with the singlet-singlet state,
\begin{equation}
\vec 8_{1,2} \cdot \vec 8_{3,4}= {1 \over 2 \sqrt{2} } \left (P_{13}-{1 \over 3}\right)
1_{1,2} \ 1 _{3,4} \ ,
\label{octet}
\end{equation}
and there is only one anti-symmetrised state,
\begin{eqnarray}
(1-P_{13})(1-P_{24}) \,
1_{1,2} \, 1 _{3,4} = 
\nonumber \\
 {1\over \sqrt{2}} 
(1-P_{13})(1-P_{24})  \,
\vec 8_{1,2} \cdot \vec 8_{3,4}  \ ,
\end{eqnarray}
where $(1-P_{13})(1-P_{24})$ is the quark and antiquark anti-symmetriser.
Therefore the multiquarks can be
described with an anti-symmetrised basis of baryons and mesons.
The multiquark may only be a bound state, or a narrow resonance,
if these baryons and mesons are sufficiently attracted. Thus the
problem of multiquark stability can be technically reduced to the 
problem of the binding of baryons or mesons. 
%
%
\begin{figure}[t]
\epsfig{file=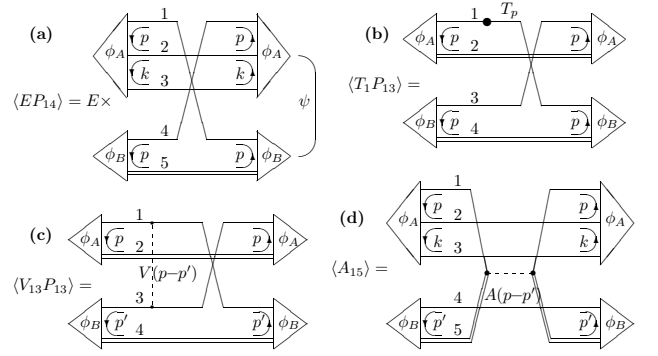,width=8.5cm} \caption{Examples of
RGM overlaps are depicted, in (a) the norm overlap for the meson-baryon
interaction, in (b) a kinetic overlap the meson-meson interaction,
in (c) an interaction overlap the meson-meson interaction, in (d)
the annihilation overlap for the meson-baryon interaction.}
\label{RGM overlaps}
\end{figure}

\par
The RGM produces both
the energy of a multiquark state and the effective hadron-hadron
interaction.  
The energy of the multiquark is computed with the matrix elements
of the hamiltonian (\ref{Hamiltonian}). The wave functions of quarks are
arranged in anti-symmetrised overlaps of simple colour singlet
hadrons. As an example, in Fig. \ref{RGM coordinates}, 
a tetraquark system is arranged in a pair of mesons $A$ and
$B$.
Once the internal energies $E_A$ and $E_B$ of the two hadronic
clusters are accounted, 
\begin{equation}
{\langle \phi_b \phi_a | H \sum_p (-1)^p P |\phi_a \phi_b \rangle
\over
\langle \phi_b \phi_a | \sum_p (-1)^p P |\phi_a \phi_b \rangle
} = E_a+E_b +V_{a \, b} \ ,
\end{equation}
where $\sum_p (-1)^p P$ is the anti-symmetrizer, the remaining energy of the meson-baryon
or meson-meson system is computed with the overlap of the
inter-cluster microscopic potentials,
\begin{eqnarray}
V_{\text{bar } A \atop \text{mes } B}
&=& \langle \phi_B \, \phi_A |
-( V_{14}+V_{15}+2V_{24}+2V_{25} )3 P_{14}
\nonumber \\
&& +3A_{15} | \phi_A \phi_B \rangle /
\langle \phi_B \, \phi_A | 1- 3 P_{14} | \phi_A \phi_B \rangle
\nonumber \\
V_{\text{mes } A \atop \text{mes } B}
&=& \langle \phi_B \,
\phi_A | (1+P_{AB})[ -( V_{13}+V_{23}+V_{14}+V_{24})  
\nonumber \\
&& \times P_{13} +A_{23}+A_{14} ]| \phi_A \phi_B \rangle 
\nonumber \\
&& / \langle \phi_B \, \phi_A | 
(1+P_{AB})(1-  P_{13}) | \phi_A \phi_B \rangle
\ ,
\label{overlap kernel} 
\end{eqnarray}
where $P_{ij}$ stands for the exchange of particle $i$ with
particle $j$, see Fig. \ref{RGM overlaps}.

\section{A criterion for binding}

In what concerns the annihilation potential, it only occurs in
non-exotic channels. Then it is clear from eqs. (\ref{sum rules}) 
that the annihilation potential provides an attractive (negative) 
overlap. This confirms that the hard core can be attractive for 
non-exotic channels where annihilation occurs.

\par
In what concerns the quark-quark(antiquark) potential,
it may also contribute to exotic channels.
Because the potential $V_{ij}$ is assumed to be proportional
to the colour dependent 
${ \vec \lambda_i \over 2} \cdot { \vec \lambda_j \over 2}$
it is clear that it can only contribute together with an exchange 
interaction, which provides a color octet, see eq. (\ref{octet}). 
Moreover the spin independent part of the interaction
vanishes. For instance in the meson-meson overlap of eq.
(\ref{overlap kernel}) the overlap of
${ \vec \lambda_1 \over 2} \cdot { \vec \lambda_3 \over 2}+
{\vec \lambda_2 \over 2} \cdot { \vec \lambda_4 \over 2}$
essentially cancels with the overlap of
${ \vec \lambda_1 \over 2} \cdot { \vec \lambda_4 \over 2}+
{\vec \lambda_2 \over 2} \cdot { \vec \lambda_3 \over 2}$.
The only potential which may contribute is the hyperfine potential, 
proportional to ${ \vec \lambda_i \over 2} \cdot { \vec \lambda_j \over 2}
\ \vec S_i  \cdot  \vec S_j $. In the present case where the kaon is a spin 
singlet, the minus phase from colour is inverted by a minus phase 
from spin. I find that the total colour and spin matrix element is a hyperfine 
splitting,
\begin{equation}
\langle P_{13}(V_{13}+V_{23}+V_{14}+V_{24})\rangle =4 {2 \over 3}(M_\Delta-M_N) \ .
\end{equation}
The flavour trace is quite simple and the spatial integral 
provides a geometrical overlap. All the corresponding factors are positive,
an therefore the quark-quark interaction results in a repulsive interaction.

\par
These results are
independent of the particular quark model that one chooses to
consider, providing it is chiral invariant.
Therefore two opposite classes of diagrams exist. The exchange
diagrams produce a repulsive interaction, which turns out to
be proportional to the hyperfine quark-quark(antiquark) interaction.
The annihilation diagrams produce an attractive interaction,
which turns out to be proportional to the spin-independent
quark-quark(antiquark) interaction.
I arrive at the attraction/repulsion criterion for the short
range hadron-hadron interaction, 
\\
- {\em whenever the two interacting hadrons have quarks (or antiquarks)
with a common flavour, the repulsion is increased,
\\
- when the two interacting hadrons have a quark and an
antiquark with the same flavour, the attraction 
is enhanced }.

\par
This paper is dedicated to the class of resonances which can be 
understood as a S=-1 kaon $s \bar u$ or $s \bar d$ strongly trapped by a s-wave hadron. 
The criterion shows that all hadrons with an antiquark $\bar u $ or $\bar d$ 
or with a a quark $s$ allow the exchange overlap $\langle P_{13}\rangle$, and this
would certainly contribute to repulsion
\cite{Bicudo,Bender,Barnes}. 
Although systems with both a short range repulsion and a short range attraction 
may still bind 
\cite{Bicudo5},
I specialize here in systems where clearly there is no hard core repulsion.
In what concerns attraction, a quark $u$ or $d$ is needed in the s-wave partner of the kaon, 
in order to produce annihilation. Therefore the isospin of the partner of the kaon needs to 
be close to the opposite isospin of the kaon. 
This excludes the mesons $\eta, \, \eta' \, \omega, \, \phi \cdots$. 
Moreover I also specialise in systems which may achieve a remarkable stability,
where the kaon partner is a hadronic resonance with a very narrow width.
This not only essentially restricts the kaon partner to s-wave mesons and baryons,
it also excludes the meson $\rho$ and the baryon $\Delta$ because they are very wide, 
due to the decay with a pion production.
The pion is also excluded because it is too light to bind to the kaon, all 
that one may get is a very broad resonance, the kappa resonance \cite{Rupp1}, which has
been recently confirmed by the scientific community. 
Therefore the hadrons which are best candidates to strongly bind the Kaons
$s \bar l$ are the s-wave hadrons with flavor 
$l\bar s , \, l \bar c, \, \l \bar b, \, lll, \, llc, \, llb , \, lcc, \, lcb, \, lbb$. 
This is expected to result in the $f_0(980)$ 
\cite{Rupp1}, 
the $D_s(2320$ 
\cite{Rupp2} 
the $D_s(2463)$
\cite{Babar}, 
the $\Lambda(1405)$ and several other resonances.

\section{The binding energy}

A convenient approximation for the meson and baryon s-wave wave-functions is 
the harmonic oscillator goundstate wave-function,
\begin{equation}
\phi_{000}^\alpha(p_\rho) =  {\cal N_\alpha}^{-1}
\exp\left({ - {{p_\rho}^2 \over 2 \alpha ^2}}\right) \ , \ \
{\cal N_\alpha} = \left({ \alpha \over 2 \sqrt{\pi}}\right)^{3 \over 2} \ ,
\label{basis}
\end{equation}
where, in the case of vanishing external momenta $p_A$ and $p_B$,
the momentum integral in eq. (\ref{overlap kernel}) is simply ${\cal
N_\alpha}^{-2}$. The coordinates of the incoming $\phi_A\phi_B$
functions are illustrated in Fig. \ref{RGM coordinates}, while the
coordinates of the outgoing $\phi_A^\dagger\phi_B^\dagger$ have
the quark 1 and 3 exchanged. I summarise 
\cite{Bicudo,Bicudo1,Bicudo2} the effective potentials computed
for the different channels,
\begin{eqnarray}
V_{K-K}&=& (2+\sigma){1\over 6} \langle A \rangle \,  {\cal N_\alpha}^{-2} \ ,
\nonumber \\
V_{K-D}&=& 2 {1 \over 6} \langle A \rangle \, {\cal N_\alpha}^{-2} \ , 
\nonumber \\
V_{K-D^*}&=& 2 {1 \over 6} \langle A \rangle \, {\cal N_\alpha}^{-2} \ , 
\nonumber \\
V_{K-B}&=& 2 {1 \over 6} \langle A \rangle \, {\cal N_\alpha}^{-2} \ ,
\nonumber \\
V_{K-B^*}&=& 2 {1 \over 6} \langle A \rangle \, {\cal N_\alpha}^{-2} \ ,
\nonumber \\
V_{K-N}&=& 3 {1 \over 6} \langle A \rangle \, {\cal N_\alpha}^{-2} \ ,
\nonumber \\
V_{K-\Sigma_c}&=& {7 \over 3} {1 \over 6} \langle A \rangle \, {\cal N_\alpha}^{-2} \ ,
\nonumber \\
V_{K-\Sigma_b}&=& {7 \over 3} {1 \over 6} \langle A \rangle \, {\cal N_\alpha}^{-2} \ ,
\nonumber \\
V_{K-\Xi_{cc}}&=& 2 {1 \over 6} \langle A \rangle \, {\cal N_\alpha}^{-2} \ ,
\nonumber \\
V_{K-\Xi_{cb}}&=& 2 {1 \over 6} \langle A \rangle \, {\cal N_\alpha}^{-2} \ ,
\nonumber \\
V_{K-\Xi_{bb}}&=& 2 {1 \over 6} \langle A \rangle \, {\cal N_\alpha}^{-2} \ ,
\label{zero p}
\end{eqnarray}
where the colour and spin factors contribute respectively with $1/3$ and $1/2$, 
$\langle A \rangle$ is of the order of 430 MeV and the geometrical factor is  
$ {\cal N_\alpha}^{-2}$. 
The remaining factor is the flavour factor. The parameter $\alpha$
is the only one that is model dependent, and it will be 
determined with a fit of experimental binding energies.
The estimation of $\alpha$ is an important by-product of this
method because the hadronic size can not be estimated directly by
the hadronic charge radius which is masked by the vector meson
dominance.
%
%
%
\begin{figure}[t]
\begin{picture}(100,150)(0,0)
\put(-70,0){\epsfig{file=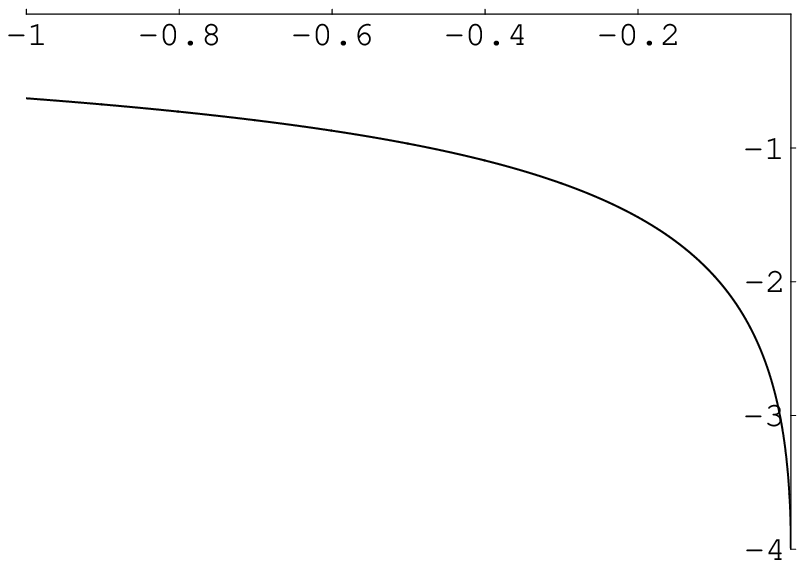,width=6.5cm}}
\put(0,140){$ E $}
\put(30,40){$g_0^1(E,1) $}
\end{picture}
\caption{ The matrix element of the Green function
$g_0(E,\mu,\alpha)$ are plotted in dimentionless units of 
$\mu=\alpha=1$. The general case scales as $g_0^\alpha(E,\mu)= 
{\mu  \over \alpha^2} g_0^1({\mu  \over \alpha^2}E ,1)$. }
\label{g0 of energy}
\end{figure}

\par
To study binding one has to proceed to the finite momentum case. 
Then the effective potentials
in eq. (\ref{zero p}) turn out to be multiplied by the gaussian
separable factor, $\exp\left[- \frac{{p_\lambda}^2}{2\beta^2
}\right] \int \frac{d^3 p'_\lambda}{(2 \pi)^3} \exp\left[- \frac{
{p'_\lambda}^2}{2 \beta^2}\right]$. In the exotic 
channels with exchange diagrams only, this result can be prooved
\cite{Ribeiro3}, 
and moreover the new parameter $\beta=\alpha$. This occurs because the overlaps
decrease when the relative momentum of the hadrons $A$ and $B$
increases. In the non-exotic channels with annihilation diagrams the
present state of the art of the RGM does not allow a precise
determination of the finite momentum overlap. Nevertheless, it is expected that 
eventually the overlap decreases due to the geometrical wave-function overlap 
in momentum space. Here the precise determination of $\beta$ does not affect
the results. I order to reduce the number of parameters, and for simplicity, 
in this paper I assume that $\beta \simeq \alpha$. 
For other approaches that also lead to a separable potential, see for instance
Ref. \cite{Cannata}.
This parametrisation of the Schr\"odinger equation in a separable potential,
\begin{eqnarray}
&& \left[ (E-T_A-T_B)(1+n |\phi^\alpha> <\phi^\alpha|) \right.
\nonumber \\
&& \ \ \left. +v|\phi^\alpha> <\phi^\alpha| \right] |\psi_\lambda>=0 \ ,
\end{eqnarray}
enables the use of standard techniques
\cite{Bicudo} 
to exactly compute the scattering $T$ matrix,
\begin{eqnarray}
T&=&|\phi^\alpha> 
{1 \over 1- {v \over 1-n} g_0^\alpha} <\phi^\alpha| \ ,
\nonumber \\
g_0^\alpha(E,\mu)&=&\langle \phi^\alpha| { 1 \over E- {p^2 \over 2 \mu }
+ i \epsilon }|\phi^\alpha \rangle  \ .
\end{eqnarray}
The binding occurs when the $T$ matrix has a pole for
a negative relative energy,
\begin{equation}
1-{v \over 1-n} g_0^{\alpha}(E,\mu)=0 .
\end{equation}
In Fig. \ref{g0 of energy} the function $G_0$ is plotted.
$G_0$  is real for a negative relative energy $E$ and is 
complex for positive $E$. The binding only occurs if,
\begin{equation}
- 4 \mu v \ge  \alpha^2 \ .
\end{equation}
Using Fig. \ref{g0 of energy} it is possible to determine the 
parameter $\alpha$ which reproduces the experimentally measured 
binding energy of the $D_s(2320)$. With a binding energy of $46$ MeV,
the corresponding parameter $\alpha$ is 285 MeV. This corresponds
to a radius of $0.7$ Fm for the nucleon.

\section{Results and conclusion}

I now compute the binding energies of the kaonic strongly bound 
molecules, or equivalently strange multiquarks. These multiquarks
are divided into two different families which are respectively
coupled and decoupled to pionic channels. The computations are
straightforward and the results are displayed in Tables 
\ref{missing couple channels} and \ref{binding energies}.

\subsection{A family of very narrow multiquark resonances}

\par
%
%
\begin{table}[t]
\begin{tabular}{c|cccccc}
channel                   
& $\mu_{exp}$  & $v_{th}$&$\alpha=\beta$& $B_{th}$& $B_{exp}$ \\
\hline 
$ D_s(2317) = {K^- {\bar D}^0+{\bar K}^0 D^- \over \sqrt{2}  }   $ 
& 392 & -143 &  285 &  {\em 46}& 46
 \\
$  D_s(2457) = {K^- {\bar D}^{*0}+{\bar K}^0 D^{*-} \over \sqrt{2} }    $ 
& 398 & -143 &  285 & 47 & 46
 \\
$ B_s = { K^- {\bar B}^0+{\bar K}^0 B^- \over \sqrt{2}  }   $ 
& 453 & -143 &  285 &  55 & -
 \\
$ B_s = { K^- {\bar B}^{*0}+{\bar K}^0 B^{*-} \over \sqrt{2}  }   $ 
& 454 & -143 &  285 & 55 & -
 \\
$  \Omega^+_{cc} = { K^- \Xi^{++}_{cc} + {\bar K}^0 \Xi^{+}_{cc}  \over \sqrt{2}  }    $ 
& 442 & -143 &  285 & 53 & -
 \\
$  \Omega^0_{cb} = { K^- \Xi^{+}_{cb}  + {\bar K}^0 \Xi^{0}_{cb}  \over \sqrt{2}  }    $ 
& 466 & -143 &  285 & 56 & -
 \\
$  \Omega^-_{bb} = { K^- \Xi^{0}_{bb}  + {\bar K}^0 \Xi^{-}_{bb}  \over \sqrt{2}  }    $ 
& 475 & -143 &  285 & 58 & -
 \\
\hline
\end{tabular}
\caption{ 
This table summarises the parameters $\mu , \, v \,
,\alpha \, , \beta$  and binding energies $B$  (in MeV)
\cite{PDB}
for the channels closed to pion decay.
The italic binding energy $B_{th}$  of the $D_s(1327)$ is fitted from experiment.
}
\label{missing couple channels}
\end{table}
The $D_s^{(0+)}$ has a lower mass by a hundred or more MeV than the 
expected $^3P_0$ excitation of the $D_s^{(0-)}(1968)$. In order to
conserve the angular momentum and parity, it can only decay
to a $D_s^{(1-)}(2112)$ with the creation of a pion, in a p-wave,
with a very low energy which is suppressed by an Adler zero
\cite{Bicudo1}. 
More importantly, this decay mode violates isospin conservation,
and therefore it is quite suppressed.
The same isolation from other strong hadronic channels is common to
other multiquarks, with a quark $s$, one or more heavy quarks and
the isospin zero combination $ u\bar u + d \bar d$.
I find that the $D_s^{(0+)}, \, D_s^{(1+)}, \, B_s^{(0+)}, \, B_s^{(1+)}$
belong to the same class of tetraquark hadronic resonances. This
class is equivalent to the picture of a kaon trapped by a
s-wave meson with a short range attraction. 
In what concerns pentaquarks, where the heavy antiquark in the
trapping meson is replaced by a pair of heavy quarks, the 
results of this paper predict that there is a similar binding with the 
quantum numbers of the $\Omega_{cc}$, $\Omega_{cb}$, and $\Omega_{bb}$. 
I conclude that the tetraquarks and pentaquarks 
$D_s^{(0+)}, \, D_s^{(1+)}, \, B_s^{(0+)}, \, B_s^{(1+)}\, 
, \Omega_{cc}, \, \Omega_{cb}$ and $\Omega_{bb}$ belong
to the same family of very narrow, non-exotic multiquarks. 
The corresponding binding energies of the kaon-hadron system
are shown in Table \ref{missing couple channels}.

\subsection{A second family, coupled to pionic channels}

\par
A second family of tetraquarks and pentaquarks may exist, where the
coupling to pionic channels does not violate the conservation of isospin.
The only suppression that one may expect comes from the Adler Zero.
The only tetraquark that we select in this class is the $I=0$ $f_0$,
which is coupled to the $\pi-\pi$ channel. 
A possible existing candidate is the $f_0(980)$, although it is not 
excluded that it is a simple $q \bar q$ meson
\cite{Bicudo4}.
In what concerns pentaquarks, when the baryon (that traps the
kaon) has two or more light quarks, the corresponding pentaquark
is again coupled to pionic channels. For instance the kaon can be
trapped by a nucleon to produce a pentaquark $\Lambda$, 
and this is coupled to the s-wave $\Sigma_s-\pi$ channel, with
the same isospin. A possible candidate to this state is the
$\Lambda(1405)$.
Similarly the kaon can be trapped by the $\Sigma_c$ or the $\Sigma_b$
baryons to produce pentaquarks with the quantum
numbers of negative parity $\Xi_c^+$, $\Xi_c^0$, $\Xi_b^0$, 
$\Xi_b^-$. These channels are respectively coupled to the 
$\pi-\Xi_c$ and $\pi-\Xi_b$ channels where the $\Xi_c$ and $\Xi_b$ have 
a positive parity. Nevertheless I compute the binding energies of this 
class of multiquarks, ignoring the effect of the coupled pionic channel. 
The results are displayed in Table \ref{binding energies}.
I find that the binding energy may be excessively large in the $f_0$ and
$\Lambda$ channel. 
This suggests that there is mixing between the two coupled
multiquark states, and that this pushes the mass of the $f_0$ and
$\Lambda$ up to the higher experimental results. 
For instance it been advocated by Oset and collaborators
\cite{Oset}
that the $\Lambda(1405)$ results from the 
mixing of a singlet-singlet state with an octet-octet state.

%
%
\begin{table}[t]
\begin{tabular}{c|cccccc}
channel                   
& $\mu_{exp}$  & $v_{th}$&$\alpha=\beta$& $B_{th}$& $B_{exp}$ \\
\hline 
$ f_0(980)= {K^- K^+ + {\bar K}^0 K^0 \over \sqrt{2} } $ 
& 248 & -215&  285 &  65 & 12 $\pm$ 10
\\
$  \Lambda(1405)={ K^- p + {\bar K}^0 n\over \sqrt{2}  }    $ 
& 325 & -215 &  285 & 88 & 30 $\pm$ 4 
 \\
$  \Xi^+_c={ K^- \Sigma^{++}_c + 2 {\bar K}^0 \Sigma^{+}_c \over \sqrt{3}  }    $ 
& 412 & -167 &  285 & 68 & -
 \\
$  \Xi^0_c={ 2 K^- \Sigma^{+}_c + {\bar K}^0 \Sigma^{0}_c \over \sqrt{3}  }    $ 
& 412 & -167 &  285 & 68 & -
 \\
$  \Xi^0_b={ K^- \Sigma^{+}_b + 2{\bar K}^0 \Sigma^{0}_b \over \sqrt{3}  }    $ 
& 456 & -167 &  285 & 75 & -
 \\
$  \Xi^-_b={ 2K^- \Sigma^{0}_b + {\bar K}^0 \Sigma^{-}_b \over \sqrt{3}  }    $ 
& 456 & -167 &  285 & 75 & -
 \\
\hline
\end{tabular}
\caption{ This table summarises the parameters $\mu , \, v \,
,\alpha \, , \beta$ 
 and binding energies $B$  (in MeV)
\cite{PDB} 
for the channels open to pion decay.
}
\label{binding energies}
\end{table}

\subsection{Comparing multiquarks with molecules and with chiral excitations}

\par
The formalism used here is also convenient to address the question,
are these new hadrons multiquarks,
or molecules composed of standard hadrons, 
or chiral images of standard hadrons?

\par
I find that, in a microscopic and chiral invariant calculation,
the new hadrons appear as tetraquarks, or as pentaquarks.
In a macroscopic hadron-hadron interaction perspective, the resulting 
mechanism which provides the binding is the short range strong attraction 
of hadrons. Technically is is not possible distinguish a microscopic
multiquark from a macroscopic strongly bound s-wave molecule, because the
hadrons totaly overlap. The purely molecular pattern only appears in different 
cases, when there is some repulsion which produces the clustering of quarks,
and this may happen in the ''pentaquark'' family of exotics
\cite{Bicudo5}.
Nevertheless the molecular perspective is convenient to estimate the mass of 
the multiquark because the binding energies are not very large, and the mass 
is essentially the sum of the kaon plus hadron mass, with a relatively small 
binding energy.

\par
The chiral excitation perspective,
present in the Chiral Soliton Model
\cite{Nowak}
or in the Chiral Lagrangian,
\cite{Lutz}
may also be eventually equivalent to 
the multiquark, or to the two-hadron narrow bound molecule. 
Suppose that a flavour singlet quark-antiquark pair $u \bar u +d \bar d$ or 
$s \bar s$ is created in a given hadron $H$. When the resulting multiquark 
$H'$ remains bound, we have a state with an opposite parity to the original
$H$. The reversed parity occurs due to the intrinsic parity of fermions and 
anti-fermions.
Moreover, because s-waves are expected to have the lowest mass,
this is equivalent to adding a pseudoscalar meson to the original $H$. 
The finite masses of the Kaon and of the pion prevent the existence of a 
large number of pseudoscalar mesons in the multiquark family of the 
$D_s(2317)$. In this family the mass shift is of the order of the kaon mass
minus the binding energy minus the strange-light quark mass difference
$M_K-B-(M_s-M_u)$, and this results in a mass shift close to 350 MeV.
The same mass shift is expected in the whole family of the $D_s(2317)$.
Thus the multiquarks $H'$ studied in this paper can be regarded as 
the chiral partners of the original $H$, in agreement with 
Nowak, Rho and Zahed. Nevertheless the standard $^3 P _0$ quark-antiquark
$D_s$ meson remains a chiral partner of the  $^1 S _0$ one, where the
mass shift is a result of the angular momentum and spin excitations. In
the present framework both the quark-antiquark creation and the angular momentum
excitation chiral excitations can be understood.  

\subsection{Outlook}

\par
The presented results only depend on the hadronic size $\alpha$ and not on the 
details of the quark-quark interactions, because the assumptions are quite simple.
I also assumed that the inverse radius parameters $\alpha$ and $\beta$ are
identical for all channels, although small channel dependences are expected. 
Moreover I neglected the meson exchange interactions because they are expected
to be smaller than the hard core interaction. In the same way I did not consider
the s-channel coupling to a single meson or baryon. More importantly, in the second 
family of multiquarks, the coupling to pionic channels was neglected although this
coupling conserves isospin. It would be particularly interesting to study these neglected 
effects in the second family studied in this paper, in order to check if they correct the 
computed binding which seems too strong. Because the research in this direction will be 
model dependent, this will be done elsewhere. 

\par
The short range attraction studied in this paper is quite general and can also be 
applied to the different candidates to multiquarks which have been discovered quite 
recently
\cite{Nakano,Barmin,Stepanyan,Barth,Alt,Belle2,Bai}. 
This will also be applied elsewhere.

\acknowledgments
I thank Emilio Ribeiro for discussions on the
RGM, and George Rupp for discussions on
hadronic resonances. I am very grateful to Gon\c{c}alo Marques for 
discussions on the family of hadronic molecules, and on the algebraic
computations of this paper. 

%

\end{document}